\documentclass[aps,prb, twocolumn, groupedaddress]{revtex4}

\usepackage[T1]{fontenc}							
\usepackage[latin1]{inputenc}	
\usepackage{graphicx}
\usepackage{amsmath}
\usepackage{amssymb}
\newcommand{\deriv}[2]{\frac{\partial^{#2}}{\partial #1^{#2}}}	
\renewcommand{\vec}[1]{\ensuremath{\textbf{#1}}}
\begin{document}
%
\title{Image Dipoles Approach to the Local Field Enhancement\\in Nanostructured Ag--Au Hybrid Devices}
\author{Christin David}
\email[corresponding author: ]{chdavid@itp.tu-berlin.de}
\author{Marten Richter}
\author{Andreas Knorr}
\affiliation{Technische Universität Berlin, Institut für Theoretische Physik, Nichtlineare Optik und Quantenelektronik, Hardenbergstraße 36, D-10623 Berlin, Germany}
\author{Inez M. Weidinger}
\author{Peter Hildebrandt}
\affiliation{Technische Universität Berlin, Institut für Chemie, Hardenbergstraße 36, D-10623 Berlin, Germany}
\date{\today}
\begin{abstract}
 We have investigated the plasmonic enhancement of the radiation field at various nanostructured multilayer devices, that may be applied in surface enhanced \textsc{Raman} spectroscopy. We apply an image dipole method to describe the effect of surface morphology on the field enhancement in a quasistatic limit. In particular, we compare the performance of a nanostructured silver surface and a layered silver--gold hybrid device. 
 It is found that localized surface plasmon states (LSP) provide a high field enhancement in silver--gold hybrid devices, where symmetry breaking due to surface--defects is a supporting factor.
 These results are compared to those obtained for multi--shell nanoparticles of spherical symmetry. Calculated enhancement factors are discussed on the background of recent experimental data.
\end{abstract}


\maketitle

\section{Introduction} 
Many spectroscopic methods studying molecular structures and processes on nanoscopic surfaces exploit the enhancement of the radiation field, resulting from the distinct local properties of the light--matter interaction.\cite{Ohtsu,NFOSPP} Surface enhancement is associated with localized surface plasmon\cite{nature,Ozbay2006} states (LSP), i.~e. the excitation of collective oscillations of the free electron gas inside rough metallic surfaces.
Here, typical effects of surface enhancement are limited to the proximity of a surface.\\
A typical example for surface enhanced spectroscopy is surface enhanced (resonant) \textsc{Raman} spectroscopy [SE(R)RS],\cite{Moskovits} where laser light of amplitude $\vec E$, that is locally amplified in the vicinity of a rough surface, is scattered by molecules attached to the surface. 
The related \textsc{Raman} scattered light is also enhanced and, given that the magnitude of the field enhancement is comparable at the frequencies of the incident and the scattered light, the total enhancement of the \textsc{Raman} intensity is proportional to the fourth power\cite{Pendry} of the enhancement of $\vec E$. However, the field enhancement factor decreases rapidly with the distance $d$ to the active surface, since the amplified electromagnetic field decays with the third power of this distance.\cite{Jackson}\\
%
%
\begin{figure}[ht!]
\centering
\includegraphics[scale=1]{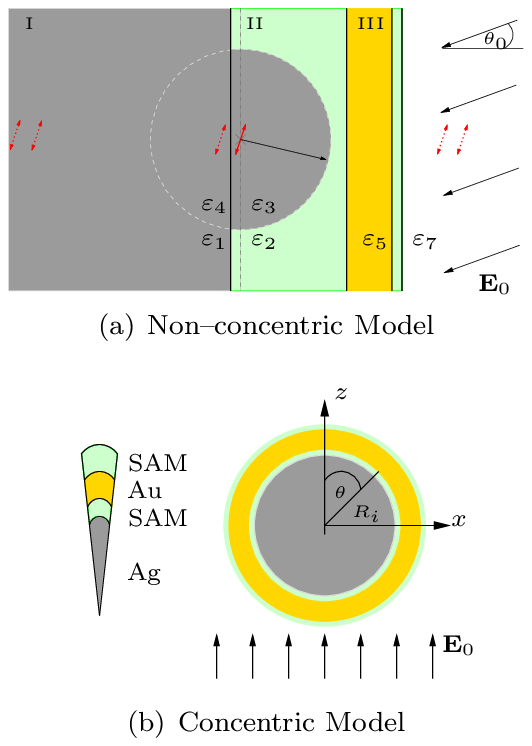}
\caption{(color online) Schematic presentation of the hybrid device geometries for the models used in this work. The device consists of a Ag core, a self--assembled monolayer (SAM), a Au layer and a second SAM. The Ag--only device is obtained from the hybrid by omitting the Au and outer SAM layer. The incoming electric field is assumed to be homogeneous in the near--field region.
a) In the non--concentric model a dipole is placed in the center of a spherical region (dipole approximation of spherical nanoparticle response). The different regions are modelled such that a Ag substrate, with a hemisphere on top modelling the surface roughness, is covered by other layers.
 For the description, the original dipole has to be mirrored at each interface and due to symmetry breaking an infinite number of image dipoles arises.
b) In the concentric model the metallic core is coated with several spherical layers. In this fully spherical, high symmetry model, only the dipole mode contributes and no image dipoles arise.
} \label{FIG:construct}
\end{figure}
The resultant enormous sensitivity of this surface enhanced spectroscopic method allows analysing molecular species at metallic surfaces and monitoring their interfacial processes. Thus, such techniques have gained increasing importance in various field of fundamental and applied sciences such as analytical chemistry, catalysis, biophysics, and material sciences.\cite{BOOK_SERS2006, BOOK_Hildebrandt2008} A drawback of this technique is the restriction to coinage metals as only for Au, Ag, and Cu a significant enhancement is obtained upon excitation in the visible (VIS) and near--infrared (NIR) spectral region. Among them, Ag exhibits by far the best optical performance in the VIS/NIR. 
This metal, however, displays a poor chemical and electrochemical stability such that its importance for analytical and specifically bioanalytical applications of SER spectroscopy is limited. For this reason and also for expanding the applicability of this method to other metals, layered hybrid systems have been designed. These devices are based on a nanoscopically rough massive Ag support, coated by a thin dielectric film and an outer Au layer.\cite{Wacker01} It was shown that the wavelength--dependence of the enhancement factor for the \textsc{Raman} scattering of molecules immobilised at the outer Au layer is essentially controlled by the optical properties of the Ag support.\cite{Hildebrandt08} 
Moreover, the magnitude of the enhancement was comparable to that for molecules directly adsorbed on Ag\cite{0_Feng08} despite the separation of ca. $20$ nm from the active surface. These results point to the possibility that optical and chemical properties of different metals can be advantageously combined, thereby paving the avenue for novel applications of SER spectroscopy.\\
The de--novo design of such devices and the optimisation of the optical properties would strongly profit from a deeper understanding of the underlying enhancement mechanism which is the central objective of the present work. Here, we have developed a theoretical approach to describe the field enhancement at model systems that mimic the nanoscopically rough Ag and Ag--Au hybrid electrodes employed in experimental studies.  In these model systems, the roughness of the real electrode surface is approximated by spherical or semispherical geometries, denoted as concentric and non--concentric model, respectively (Fig. \ref{FIG:construct}).  The various components of the layered devices include Ag, self-assembled monolayers (SAMs) of mercaptanes, Au, and water and possess specific dielectric properties.\\
A critical aspect of the theoretical model refers to the description of the Ag surface roughness. In the non--concentric model we have considered a hemisphere on a flat surface whereas the outer Au layer is taken to be flat. This approximation is justified since the Au surface roughness in hybrid electrodes is lower than that of the Ag surface.\cite{0_Feng08} Nevertheless, this model provides a more realistic picture of the real electrodes than the concentric model. The latter represents a useful reference system for sorting out the effects of symmetry breaking that are supposed to have a large effect on local field enhancement.\cite{Halas2008}
In this low--symmetry approach, higher multipoles contribute to the field enhancement in contrast to the concentric multi--shell model, where only the dipole mode is excited\cite{Jackson} and the field distribution displays this symmetry.\\
The near--field properties of the electromagnetic field typically dominate the optical interaction in nanostructures. Both of our models are,  therefore, examined in a quasistatic limit, where the electrical field becomes longitudinal $\vec E=-\nabla\Phi$, retardation effects are excluded and the homogeneous Poisson's equation $\Delta\Phi(\vec r, \omega)=0$ is solved in frequency space for the geometries depicted in Fig. \ref{FIG:construct}.\\
The article is organized as follows: In section II we outline the dielectric response function used to describe the metal layers. Subsequently, in section III, the image dipole approach for a multilayer structure in a non--concentric geometry and the concentric model are discussed. We present the results for the devices in both models in section IV. 
\section{Dielectric function} 
An incident electric field causes polarization inside the material. The response of the system is described by its dielectric function $\varepsilon(\vec r, \omega)$. For metals this function is often modelled on the basis of the \textsc{Drude} model for the free electron gas in a single band (intraband effects\cite{ashmer}).\\
However, with increasing frequency $\omega$ the contribution of excited {d--band} electrons (occuring interband transitions) becomes important for metals as well\cite{AlvarezEtAl} and has to be included for a realistic characterization of the metallic response function. We, therefore, write for metallic regions
\begin{align}
\varepsilon(\omega)&=\varepsilon_D(\omega) + \varepsilon_{IB}(\omega), \label{eq:eps_all}
\intertext{with the \textsc{Drude} part}
\varepsilon_D(\omega)&=\varepsilon_{\infty}-\frac{\omega^2_{pl}}{\omega(\omega+{\rm i}\gamma)},	\label{SPP:Drude}
\intertext{whereby $\varepsilon_\infty$ refers to the high frequency limit and $w_{pl}$ is the plasmon frequency for bulk material. Taking into account that the free mean path is limited in small particles with radius $R$,\cite{Kreibig1969} the damping $\gamma$ can be written as a size-dependent and a constant damping part}
\gamma(R)&=A\frac{v_F}{R}+\gamma_{pl}.	\label{intro:gamma}
\end{align}
Here, $\gamma_{pl}$ and $v_F$ refer to the inverse lifetime of plasmonic oscillations and the \textsc{Fermi} velocity, respectively. The parameter
$A\; (\sim1)$ accounts for the scattering process.\cite{HoevelEtAl} An even better description of size effects can be obtained by using non--local
theories.\cite{Abajo2008_nonlocal}\\
Note, that our description does neglect the influence of spatial dispersion. As previously shown,\cite{McMahon2009}
such effects enhance spectral shifts and structures and yield corrections to field enhancement in very small metal structures (<10nm). Such effects
should be taken into account if a more quantitative description, in particular for single, individual structures is needed.\\
The interband transition part reads:\cite{Etchegoin2006,Etchegoin2007,Vial2007}
\begin{align}
\varepsilon_{IB}(\omega)&=\sum_iA_i\omega_i\left(\frac{e^{i\Phi_i}}{\omega_i-\omega-\frac{{\rm i} 2\pi c}{\gamma_i}}+\frac{e^{-i\Phi_i}}{\omega_i+\omega+\frac{{\rm i} 2\pi c}{\gamma_i}}\right). \label{eq:IB_part}
\end{align}
The sum indicates the number of interband transitions, typically ${\rm i}\in\{1,2\}$ for the considered spectral range. We denote the transition frequencies with $\omega_i$, $\gamma_i$ are the corresponding damping parameters in nm (linewidth), $\Phi_i$ denote phases and $A_i$ are dimensionless amplitudes. Eq. \eqref{eq:eps_all} is then fitted to experimental data of Ag\cite{JohnsonChristy} to determine the individual parameters in Eqs. \eqref{SPP:Drude}-\eqref{eq:IB_part}, see Table \ref{TAB:Fitting_values}. For Au, this procedure has already been employed by \textsc{Etchegoin} et al.,\cite{Etchegoin2006} using experimental data for Au reported in an earlier study.\cite{Etchegoin2007}\\
\begin{table}
 \centering
 \caption{Parameters for the dielectric functions of silver for Eq. \eqref{eq:eps_all} used to fit experimental data.\cite{JohnsonChristy}}
\begin{ruledtabular}
 \begin{tabular}{lll|cccc} 
$\varepsilon_\infty$	&		&				& 	&	& 0.825 &	\\
$\omega_{pl}$		& 		&[eV]				&	&	& 9.143 &	\\
$\gamma_{pl}$		& 		&[eV]				&	&	& 0.021 &	\\
$A_1,$			& $A_2$		&				&	& 0.313	& 	& 1.329	\\
$\omega_1,$		& $\omega_2$	& [eV]				&	& 4.080	& 	& 5.227	\\
$\Phi_1,$		& $\Phi_2$	&				&       & -1.372& 	& -0.456\\
$\gamma_1,$		& $\gamma_2$	& [nm]				&	& 0.446	& 	& 2.812	\\
$v_F$ 			& 		& $[\frac{\rm{nm}}{\rm{fs}}]$	&	&	& 1.4 	&
 \end{tabular}
\end{ruledtabular}
 \label{TAB:Fitting_values}
\end{table}
In contrast for the wavelength--dependent complex dielectric function for the metals, the SAM layers are described by a wavelength--independent
dielectric constant of $\varepsilon_{SAM}=2$.
\section{Image Dipoles Approach} 
The use of image dipoles to analytically describe the electric field distribution at spherical particles or caps in front of a substrate has been discussed in detail by \textsc{Wind} et~al..\cite{Wind} In this work, that approach is extended to a multilayer structure.\\
We simulate the roughness of the real Ag surface with a Ag hemisphere on a Ag substrate (half plane) in front of the other layers. In this non--concentric model a dipole is placed in the center of a spherical region with radius $R_{Ag}$, to model the response of the hemisphere in a dipole approximation, see Fig.~1(a). 
To obtain a correct solution, this dipole has to be mirrored at each interface. Due to symmetry breaking in the multilayer structure, an infinite number of image dipoles immediately arises when assuming at least two interfaces, as it is the case for the Ag electrodes as used in experiments.\cite{0_Feng08}\\
The distance of image dipoles from the hemisphere is increased each time they are mirrored, which allows us to reduce the following considerations to a finite number of image dipoles.\\
It is convenient to work with a reduced, i.~e. dimensionless, potential  $\psi(\vec r, \omega)=-\frac{\Phi(\vec{r}, \omega)}{E_0R_{AgAu}}$, where
$E_0$ denotes the amplitude of a spatially homogeneous incident electric field. $R_{AgAu}$ denotes the distance from the substrate (center of the
hemisphere) to the SAM--H$_2$O interface.\\
The solution of the homogeneous \textsc{Poisson}'s equation is given by an expansion in associated \textsc{Legendre} polynomials of zeroth ($P^0_j$) and first ($P^1_j$) order. The latter accounts for an additional degree of freedom (azimuthal angle $\phi$) which vanishes under full spherical symmetry.  The argument of the \textsc{Legendre} polynomials is displaced by the image dipoles (virtual) coordinates in units of $R_{AgAu}$, with the original dipole at $\vec r_{i=0}=0$.\\
In general, the multipolar reduced potential in a region $\epsilon$ (index associated with the dielectric function $\varepsilon(\omega)$ of this
region) with
image dipoles at sites $\vec r_i$ and $\eta=\cos\theta$ (cp. Fig.~1) reads
\begin{align}
\psi^\epsilon&(r,\theta,\phi)=\tau^\epsilon r\eta\cos\theta_0+\delta^\epsilon r\sqrt{1-\eta^2}
\sin\theta_0\cos\phi+\psi^{\epsilon}_0\nonumber\\\nonumber
&+ \sum_{i,j=0}^{\infty}P_j^0\bigg(\frac{r\eta-r_i}{|\vec r-\vec r_i|}\bigg) \big[\frac{A_{ij}^{\epsilon}}{|\vec r-\vec r_i|^{j+1}}+\tilde{A}_{ij}^{\epsilon}|\vec r-\vec r_i|^{j}\big]\\
&+ \cos\phi\sum_{i,j=0}^{\infty} P_j^1\bigg(\frac{r\eta-r_i}{|\vec r-\vec r_i|}\bigg) \big[\frac{B_{ij}^{\epsilon}}{|\vec r-\vec r_i|^{j+1}} + \tilde{B}_{ij}^{\epsilon} |\vec r-\vec r_i|^{j}\big],\label{eq:mult:pot}
\end{align}
where the first line describes the direct influence of the incident field with incident angle $\theta_0$. $A_{ij}^\epsilon,\tilde{A}_{ij}^\epsilon$
and $B_{ij}^\epsilon,\tilde{B}_{ij}^\epsilon$ are the multipole coefficients, that have to be determined by examining the related boundary conditions.
They describe the two lowest modes ($m=0, 1$) of the associated \textsc{Legendre} polynomials $P^m_j$. Higher modes $m>1$ are not excited by the
incident field.\\
Since the distance of image dipoles increases every time they are mirrored, it is justified to truncate the sum over index $i$ at a certain value, that depends on the number of layers. For the coated Ag surface with two planar interfaces we find ten dipoles to be sufficient, whereas for the hybrid device model we used 62 dipoles.
Similarly, the truncation limit for summing up the multipole contributions was chosen after examining the convergence of the second sum in Eq. \eqref{eq:mult:pot}.\\
For setting up the electrostatic potential, it is necessary to deduce, which image dipole enters into the potential of a given region. The real dipole, however, influences all regions. When mirroring the real dipole at an interface, it is convenient
to use the following mnemonic for the virtual dipoles: The potential of each dipole ``in front'' of a layer contributes to the potential describing that layer, whereas the potential of each dipole ``behind'' the interface does not. Thus, when an image dipole emerges from mirroring at an interface, it is not contained in the region where it is projected into. Therefore, only the layer embodying the original dipole is described by a potential, where all considered dipoles enter.
In contrast to the findings of \textsc{Wind} et al.\cite{Wind} for a single layer, analytic expressions for a multilayer structure can only be derived for a hemisphere ($r_0\rightarrow0$) and not for arbitrary distances $r_0$ of the real dipole to the substrate layer. For a layered device however, it is necessary to consider $r_0\neq0$ in the analysis.
\section*{Relations between Multipole Coefficients} 
When studying boundary conditions at an arbitrary planar interface $r\eta=z=z_0$, seperating the regions $\epsilon$ and $\epsilon'$
\begin{align}
 {\psi^{\epsilon}}|_{z_0}={\psi^{\epsilon'}}|_{z_0},\quad\varepsilon(\omega)\deriv{z}{}{\psi^{\epsilon}}|_{z_0}=\varepsilon'(\omega)\deriv{z}{}{\psi^{
\epsilon'}}|_{z_0}
\end{align}
we find different kinds of relationships between the multipole coefficients for an arbitrary set of dipoles $\{i\}_{i\in\mathbb{N}}$.
For instance, for multipole coefficients of planar interfaces, that do not touch the layer with the hemisphere, we obtain:
\begin{align}
A^{\epsilon}_{ij}=\frac12\left(A^{\epsilon'}_{ij}\big(1+\frac{\varepsilon'}{\varepsilon}\big)+A^{\epsilon'}_{i+1,j}\big(1-\frac{\varepsilon'}{\varepsilon}\big)\right),\\
A^{\epsilon}_{i+1,j}=\frac12\left(A^{\epsilon'}_{ij}\big(1-\frac{\varepsilon'}{\varepsilon}\big)+A^{\epsilon'}_{i+1,j}\big(1+\frac{\varepsilon'}{\varepsilon}\big)\right).
\end{align}
Here, $i+1$ denotes the lowest order image dipole to $i$. Note that in the limit of a hemisphere two such dipoles coincide, which leads to the simple relationship:
\begin{align}
 A^{\epsilon}_{ij}+A^{\epsilon}_{i+1,j}=A^{\epsilon'}_{ij}+A^{\epsilon'}_{i+1,j}. \label{eq.:simple_addition}
\end{align}
for the outer planar regions. In that case, the field distribution of an outer planar region is just given by the displaced field distribution of its neighbouring layer.\\
For planar regions touching the layers with the spherical region ($A^{\epsilon'}_{i+1,j}=0$), we find
\begin{align}
   A^{\epsilon'}_{ij}&=\frac{2\varepsilon}{\varepsilon+\varepsilon'}A^{\epsilon}_{ij},	\qquad	 A^{\epsilon}_{i+1,j}=(-1)^j\frac{\varepsilon-\varepsilon'}{\varepsilon+\varepsilon'}A^{\epsilon}_{ij},
\end{align}
using the property $P^m_j(-\eta)=(-1)^{j+m}P^m_j(\eta)$.\\
The coefficients of the regions within the hemisphere match the same way, we only have to choose the proper dielectric functions. 
Note that these two types of relations between the coefficients have equal resonance(s) for vanishing $\varepsilon(\omega)+\varepsilon'(\omega)$ and are determined by the three inner planar regions: Ag, SAM and Au (Fig. \ref{FIG:construct}).
With the relations derived above, all $A^\epsilon_{ij},\tilde A^\epsilon_{ij} $ (and $B^\epsilon_{ij}, \tilde B^\epsilon_{ij}$ with the substitution\footnote{In fact, all powers $j$ are $j+m$. The coefficients $A_{ij}^\epsilon$ are associated with $m=0$ and $B_{ij}^\epsilon$ with $m=1$. However, restricting our considerations to the plane containing all dipoles which is orthogonal to the layers, we can omit the coefficients $B_{ij}^\epsilon$ and $\tilde{B}_{ij}^\epsilon$.} $j\longrightarrow j+1$) can be expressed in terms of $A^{II,o}_{0j}$, describing the region II (see Fig.~1(a)) with the original dipole outside and $A^{II,\rm i}_{0j}$ inside the spherical particle. The choice of region II is arbitrary, but it is resonable since $A^\epsilon_{0j}$ describes the influence of the original dipole in region $\epsilon$.\\
The coefficients $A_{ij}^\epsilon$ and $\tilde{A}_{ij}^\epsilon$ are matched via the multipolar polarizability of order j:
\begin{align}
\alpha_j(\omega)=R^{2j+1}\frac{j(\varepsilon(\omega)-\varepsilon'(\omega))}{j(\varepsilon(\omega)+\varepsilon'(\omega))+\varepsilon'(\omega)}.
\end{align}
In reduced coordinates $\vec r$ we obtain:
\begin{align}
 \tilde{A}_{ij}^{\epsilon}{|\vec r-\vec r_i|_{|z_0}^{2j+1}} &=\frac{(j+1)(\varepsilon-\varepsilon')}{(\varepsilon+\varepsilon')j+\varepsilon'}A_{ij}^{\epsilon}.
\end{align}
The last boundary condition leads us to conditional equations to find the multipole coefficients $A^{II,o}_{0j}, A^{II,\rm i}_{0j}$.
\section*{Final Conditional Equations} 
In this section we derive conditional equations for the multipole coefficients.\cite{Wind} 
For this purpose, we use the remaining boundary conditions at the surface of the sphere, i. e. $r=1$, and take advantage of the properties of the associated  \textsc{Legendre} polynomials by multiplying with $P_k^0(\eta)$ and $P_k^1(\eta)\cos\phi$, and integrating over the sphere. Thus, we find separate conditional equations for the coefficients $A^{II,o}_{0j}, A^{II,\rm i}_{0j}$ and $B^{II,o}_{0j}, B^{II,\rm i}_{0j}$.
\begin{align}
\phantom{+}\int_0^{2\pi}d\phi\bigg(&\int_{-1}^{0}d\eta P^0_k(\eta)(\psi^{II,o}-\psi^{II,\rm i})|_{r=1}\nonumber\\
+&\int_{0}^1d\eta P^0_k(\eta)(\psi^{I,o}-\psi^{I,\rm i})|_{r=1}=0\bigg), \label{condEqA1}\\
 \phantom{+}\int_0^{2\pi}d\phi\cos\phi\bigg(&\int_{-1}^{0}d\eta P^1_k(\eta)(\psi^{II,o}-\psi^{II,\rm i})|_{r=1}\nonumber\\
+&\int_{0}^1d\eta P^1_k(\eta)(\psi^{I,o}-\psi^{I,\rm i})|_{r=1}\bigg)=0\label{condEqB1}
\end{align}
and for the discontinuity of the normal component
\begin{align}
 \int_0^{2\pi}d\phi&\bigg(\int_{-1}^{0}d\eta P^0_k(\eta)\deriv{r}{}(\varepsilon_1\psi^{II,o}-\varepsilon_3\psi^{II,\rm i})|_{r=1} \nonumber\\
+\int_{0}^1&d\eta P^0_k(\eta)\deriv{r}{}(\varepsilon_2\psi^{I,o}-\varepsilon_4\psi^{I,\rm i})|_{r=1}\bigg)=0,\label{condEqA2}\\
 \int_0^{2\pi}d\phi&\cos\phi\bigg(\int_{-1}^{0}d\eta P^1_k(\eta)\deriv{r}{}(\varepsilon_1\psi^{II,o}-\varepsilon_3\psi^{II,\rm i})|_{r=1}
\nonumber\\
+\int_{0}^1&d\eta P^1_k(\eta)\deriv{r}{}(\varepsilon_2\psi^{I,o}-\varepsilon_4\psi^{I,\rm i})|_{r=1}\bigg)=0.\label{condEqB2}
\end{align}
Integration over $d\phi$ and $\cos\phi d\phi$ selects the multipole coefficients $A^{II,o}_{0j}, A^{II,\rm i}_{0j}$ and $B^{II,o}_{0j}, B^{II,\rm i}_{0j}$, respectively. Note, that $\theta_0$ is the incident angle of the electric field in the $x-y$--plane.\\
Upon matching the boundary conditions, we have reduced the coefficients $A^{II,o}_{0j}, A^{II,\rm i}_{0j}$ to the original dipole $(i=0)$. In this way, we obtain frequency dependent parameters, built up by the dielectric functions of the different regions and overlap integrals of displaced associated \textsc{Legendre} polynomials. The latter can only be simplified in an analytic way for the original dipole and its lowest order image dipole. The convergence of the multipole coefficients has to be verified, i.~e. higher multipole coefficients have to be sufficiently small for truncation of the sum at $j=J$.\\
Eqs. \eqref{condEqA1}-\eqref{condEqB2} can finally be written as\cite{Wind}
\begin{align}
  \sum_j\alpha_{jk}A^{II,o}_{0j}+\sum_j\beta_{jk}A^{II,\rm i}_{0j}+\cos\theta_0\gamma_k&=0, \label{Cond01}\\
  \sum_j\iota_{jk}A^{II,o}_{0j}+\sum_j\zeta_{jk}A^{II,\rm i}_{0j}+\cos\theta_0\kappa_k&=0, \label{Cond02}
\end{align}
with frequency and geometry dependent matrix elements. Eqs. \eqref{Cond01}, \eqref{Cond02} are valid for an arbitrary set-up with a single spherical region and various surrounding planar layers. Again, for $B^{II,o}_{0j}$ and $B^{II,\rm i}_{0j}$ we find similar expressions. This set of linear equations is the starting point for a numerical analysis. Solving these equations for $J$ multipole coefficients yields the electrostatic potential and, therefore, the local field enhancement:
\begin{align}
 g_0=\left| \frac{\vec E(\vec r,\omega)}{\vec E_0} \right|^2=\left|\nabla_{(r,\theta,\phi)}\psi(\vec r,\omega)R \right|^2.
\end{align}
The enhancement factor for the \textsc{Raman} intensity, analysed in the next chapter, is then given by $g_0^2$. 
In a concentric multilayer model, i.~e. for a nanoparticle coated with several spherical shells, no image dipoles emerge and the electrostatic potential can be derived analytically. When matching the boundary conditions one even finds, that only the dipole contributions are excited.\cite{Jackson} This leads to a polarization field with radial symmetry, a constant field within the core and a symmetrically distributed total field. 
\section{Results} 
We compare the field enhancement $g_0$ as function of the wavelength for the two multilayer systems in both (non--concentric in Fig.~2(a), concentric in Fig.~2(b)) models. Here, the field enhancement by the Ag--only model is evaluated at the distance corresponding to the SAM--H$_2$O interface of the hybrid device $R_{AgAu}$. Since a real experiment collects the signal from different inhomogenities, the wavelength--dependence of $g_0$ is averaged over several layer thicknesses to simulate the response of an array of different sized hemispheres, with a \textsc{Gaussian} distribution around mean values 
$40\pm20$ nm for the radius of the hemisphere and $16\pm10$ nm for the gold layer. Interactions between the (hemi-) spherical structures are not included in this averaging procedure and we obtain the mean field enhancement of non--interacting nanostructures.
%
%
\begin{figure}[ht]
 \includegraphics[scale=1]{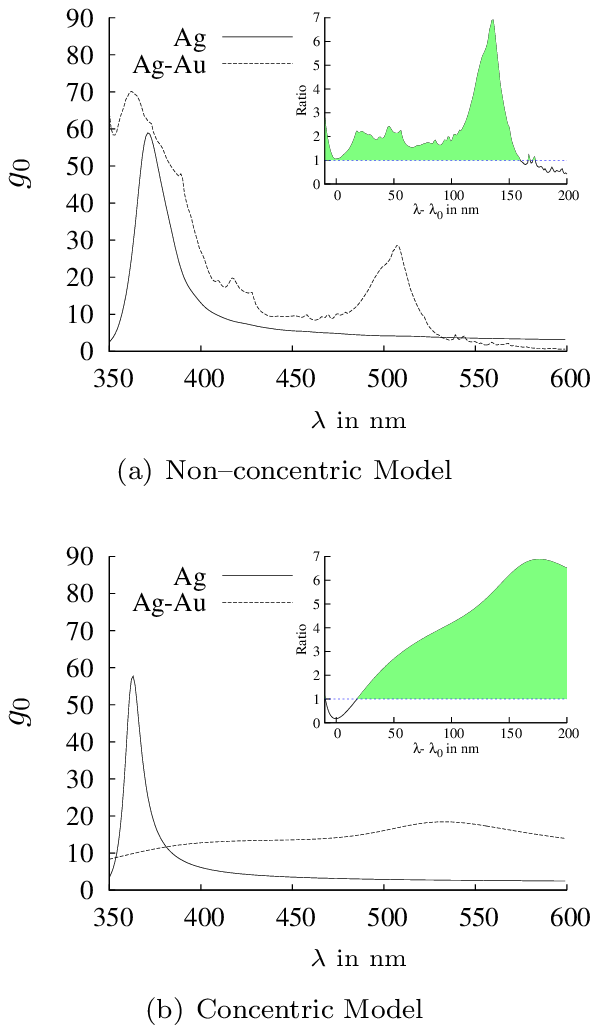}
 \caption{(color online) Wavelength--dependence of the field enhancement for both devices, evaluated at the distance of the (Au)SAM--H$_2$O interface $R_{AgAu}$. The Ag--only device exhibits a sharp plasmonic resonance, whereas the hybrid device has lower maximum values for the field enhancement intensity but reveals a wide enhancement profile. This is due to an overlap of the contributions of silver and gold resonances.
The insets show the ratio of the field enhancement of the Ag--Au hybrid compared to the Ag--only device relative to wavelength of the Ag plasmon resonance $\lambda_0$. It can thus be seen, that the hybrid device exceeds the maximum enhancement of the pure Ag system by a factor up to seven.
} \label{Fig:spectral}
\end{figure}
For both geometries, the model affords a sharp plasmonic resonance with comparable values (Fig. \ref{Fig:spectral}, solid line). Due to geometry induced effects, the wavelength of the resonance is slightly redshifted in the non--concentric model.\\
In contrast to the Ag structure, the Ag--Au hybrid device (dotted lines in Fig. \ref{Fig:spectral}) shows lower maximum enhancement 
but a distinctly broader enhancement profile with much higher values compared to the Ag device upon excitation out of the resonance in respect to the plasmonic frequency of Ag.
The overlap of broadened Ag and Au plasmonic resonances contribute to the wavelength--dependence of the field enhancement for the hybrid device. Whereas in the concentric multi--shell model (Fig.~2(b)) the wavelength--dependence seems to be almost uniform, in the non--concentric model (Fig.~2(a)) both contributions are well distinguishable and even blueshifted compared to the results of the concentric model. Obviously, the prevalence of the Ag plasmonic resonance in the non--concentric model is a consequence of the symmetry breaking due to the geometry of the Ag surface. In this low symmetry model, the field enhancement is concentrated at the edge between Ag substrate and hemisphere, see Fig.~3(a). Thus, the field enhancement is restricted to the front side facing the Au layer.
In this respect, the results are reminiscent to previous findings for a core displaced within a shell.\cite{Halas2008}
\textit{For SER spectroscopy, a spectrally broad enhancement profile is advantageous, since it allows for a comparable field enhancement both at the frequency of the incident and of the \textsc{Raman} scattered light.}\\
The present calculations predict a superior optical performance for hybrid devices as it is illustrated by the insets in Fig. \ref{Fig:spectral}. These insets show the ratios of the field enhancement of the Ag and Ag--Au geometries as a function of the frequency separation relative to the plasmon resonance of the Ag model, denoted as $\lambda_0$. Again, the results show, that already a few nanometers displaced from the Ag resonance peak the hybrid device exceeds the enhancement of the Ag--only device in both models.
Under these conditions, the Ag--Au hybrid structure provides field enhancement factors that are up to seven times higher than those achieved with the Ag--only model. \\
Comparing to the plasmonic resonance of Au nanoparticles, which is around $560$ nm, the second resonance peak in Fig.~2(a) can be attributed to the gold layer, but is blueshifted due to geometry induced effects. Due to the non--uniform enhancement profile of the Ag--Au hybrid device in the non--concentric model, the range where the hybrid device shows higher enhancement factors is smaller ($400 - 550$ nm) than in the concentric model.\\
In our non--concentric model, see Fig.~3(a), the polarization field within the gold layer is not fully anti--parallel to the incoming light, leading to higher enhancement factors within the Au layer than in the concentric model. 
On the other hand, the mean factor by which the hybrid device exceeds the enhancement of the Ag support is about two in the non--concentric model, cf. Fig.~2(a).\\
In Fig. \ref{FIG:contours} the spatial distribution of the normalized field enhancement at an excitation wavelength of $413$ nm shown as a function of the reduced coordinates for the Ag--Au hybrid device.
For the non--concentric model, the metal hemisphere is readily identified due to very low field enhancement (dark color), which is attributed to  polarization effects inside the metal reducing the incoming field. The hemisphere is part of the Ag support with its planar interface at $z=0$. 
The edges between the substrate and the hemisphere are sharper than the curvature of the sphere itself, leading to a particulary high field enhancement. The enhancement in the dielectric gap between Ag and Au is much lower in the non--concentric model than compared to the gap in the concentric model. Here the field enhancement in the dielectric spacer exceeds that at the outer Au layer.\\
%
%
\begin{figure}[ht]
\includegraphics[scale=1]{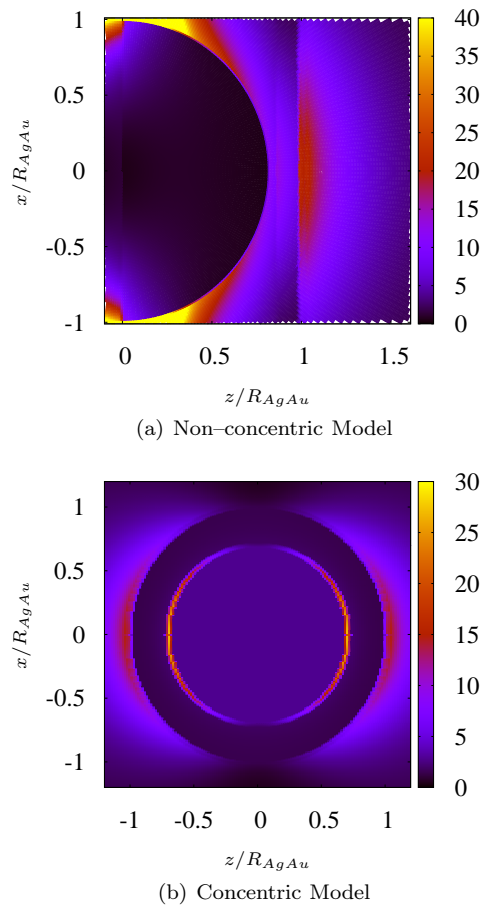}
 \caption{(color online) Full spatial distribution of the field enhancement of the hybrid device in both models. a) In the non--concentric model the polarization effects within the silver hemisphere lead to very low field intensities in the metal (black). Very high field enhancement (yellow), due to LSP states, is observed at the sharp edge between the hemisphere and the planar silver substrate. 
b) In the high symmetric model, the field enhancement is symmetrically distributed around the coated metal nanosphere. A high enhancement factor is achieved in the dielectric spacer (SAM), the shell between the two metals. Due to anti--parallel polarization, the electric field in both metals is lowered.}
 \label{FIG:contours}
\end{figure}
The field in the Au layer is also influenced by a polarization field leading to low enhancements within this layer. On top of the Au layer, however, high field enhancements are predicted. The polarization effects within the metals lead to an increase of field intensity at the surfaces to the dielectric layers. In the non--concentric system the electric field is redistributed and concentrated on the dielectric side of the interface. Therefore, symmetry breaking is an important ingredient for the high field enhancement at the surface of the devices.\\
In the concentric model, the strongest enhancement is predicted parallel to the incident light field, whereas nearly no field enhancement is calculated in perpendicular direction. This behaviour is comparable to a source of dipole radiation, reflecting the fact that only the dipole mode of the \textsc{Legendre} polynomials contributes to the solution. In contrast, the hybrid device displays a strong enhancement within the inner dielectric layer almost independent of the direction.\\
%
%
\begin{figure}[ht]
\includegraphics[scale=1]{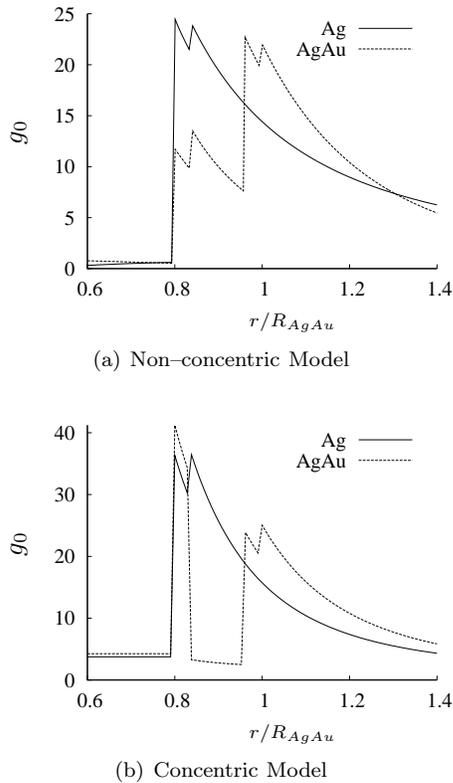}
 \caption{Spatial distribution parallel to the incident electric field vector (z--direction) through the center of the spherical particle at an excitation wavelength of $413$ nm. The thickness of the gold layer, the SAM and the Ag core radius are $6$ nm, $2$ nm and $40$ nm, respectively. Both models show low field enhancement within the Ag core due to a polarization field inside the metal. a) The non--concentric model is consistent with experimental observations\cite{0_Feng08,Hildebrandt08}, i.~e. an enhancement factor at the outer Au--SAM surface that is comparable to that at the SAM--coated Ag--only device. Within the planar Au layer the polarization field is not fully anti--parallel leading to higher field strength close to the hemisphere. b) The concentric model predicts a high excitation at the Ag core but far lower enhancements at the outer Au surface. The maximum values at the core itself are higher here, since in this model the curvature of the spherical particle is the sharpest, producing highest enhancement factors. Due to high symmetry also within the gold layer an anti--parallel polarization field is established. The strong reduction of the field intensity leads to a reduced enhancement at the Au--SAM--H$_2$O interface.
} \label{FIG:spatial}
\end{figure}
Fig. \ref{FIG:spatial} presents the distance--dependent enhancement profile for both devices at $x=0$. The Ag--only system displays a strong field enhancement at its metal-to-dielectric interface (SAM--H$_2$O), where two materials of different permittivitty are in contact.\\
The hybrid device produces a remarkable enhancement at the outer Au--SAM--H$_2$O interface. The Ag core is sufficiently excited by the incident light although a $6$nm gold layer has a transmittance of ca. $60 \%$ for excitation wavelengthes below $500$ nm.\\
At the attachment site for the target molecule in the Ag--Au hybrid material, i. e. the outer (Au)SAM surface, both geometric models predict a field enhancement for $\lambda=413$ nm that is by $63 \%$ higher than at the Ag surface at that distance.
In the concentric model, the field enhancment for the hybrid material is about one third smaller compared directly to the SAM--H$_2$O interface of the Ag surface, but of comparable values in the image dipoles approach, which agrees with the findings in recent experiments.\cite{0_Feng08}
The local field within the hemisphere is not constant as in the concentric model and, due to symmetry breaking, the field enhancement at the inner dielectric spacer is not as high between the two metals as in the concentric model.
The polarization field within the Au layer, that reduces the electric field in the metals, is fully anti--parallel in the concentric model in contrast to the image dipoles approach, where the enhancement factors within the Au layer are higher.\\
The field enhancement is limited to the dielectric region between the metal layers, the localized surface plasmons (LSP) form a gap mode.
The optical energy is then mediated through the Au layer and high field enhancement is additionally produced at the outer Au surface.\\
%
%
\begin{figure}[ht]
 \includegraphics[scale=1]{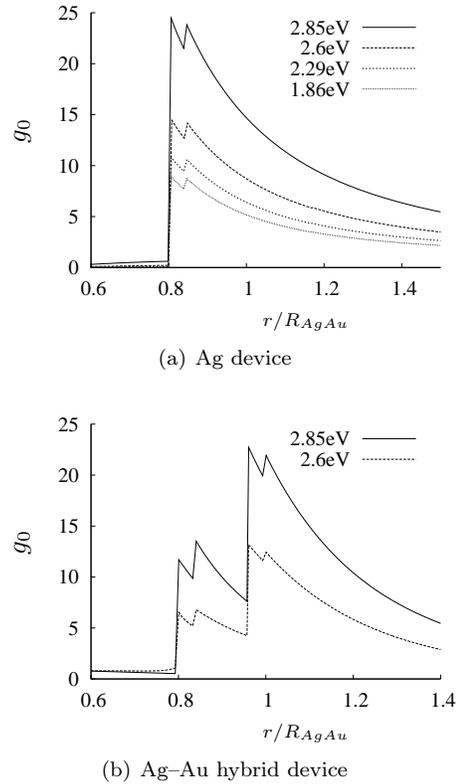}
 \caption{For the non--concentric model the spatial distribution at both devices is displayed for different excitation frequencies: 2.85~eV (ca. 413~nm), 2.6~eV (453~nm), 2.29~eV (514~nm) and 1.86~eV (634~nm). For sake of clarity, we show only two traces for the hybrid device. 
(a) SAM--coated Ag electrode in aqueous solution: Ag(R=40~nm)-SAM(d=2~nm)-H$_2$O. 
(b) Ag--Au hybrid device: Ag(40~nm)-SAM(2~nm)-Au(6~nm)-SAM(2~nm)-H$_2$O. 
Both devices reveal similar field enhancement factors at their respective SAM--H$_2$O interfaces. 
}\label{FIG:spatial_freq}
\end{figure}
Under off--resonant excitation (Fig.~\ref{Fig:spectral}), the field enhancement and thus the predicted Raman enhancement ($\sim g_0^2$) of the hybrid device at the molecule attachment site is even higher.
In Fig.~\ref{FIG:spatial_freq} the spatial field distribution of the two devices\cite{0_Feng08} is presented for different excitation frequencies.\\
%
%
\begin{figure}[ht]
\includegraphics[scale=1]{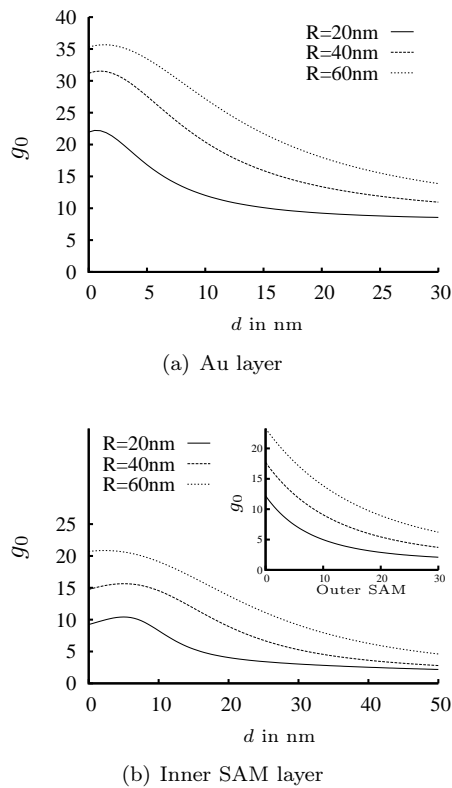}
 \caption{Field enhancement factor of the Ag--Au hybrid device as a function of the various layer thicknesses and different core sizes, evaluated for an excitation wavelength of $413$ nm.}
 \label{FIG:GapDep}
\end{figure}
For the layered hybrid device, the dependence on different layer thicknesses is particularly instructive for guiding the de--nevo design of efficient SER--active materials (Fig.~\ref{FIG:GapDep}). 
With growing Ag core size the field enhancement increases approaching a saturation for $R\approx 120$nm.\\
The enhancement gradient is affected by the thickness of the Au layer, for which optimum values are 2 to $5$~nm, for a fixed SAM thickness of $2$~nm.\\
The thickness of the inner dielectric spacer strongly affects the enhancement, due to the formation of the LSP modes between the metallic layers. For small radii of the Ag core an optimum enhancement can be found around $5$~nm. The distance between the two metal layers determines the mediation of optical energy via plasmonic excitation. It can also be concluded that two metals in direct contact still show a non-vanishing enhancement.
Then the plasmonic excitation of a coated metallic nanoparticle with an effective dielectric function can be assumed.\\
The outer dielectric layer (inset in Fig.~6(b)) shows no special features. It is comparable to the pure Ag surface with increasing outer SAM layer with equal distance dependence ($g_0^2\sim d^{-12}$).
\section{Conclusion and Outlook} 
The Ag-Au hybrid devices treated in this work display a spectrally borader enhancement profile as compared to the pure Ag systems that exhibit a single, sharp plasmonic resonance peak. The additional metallic layer in the Ag--Au hybrid devices is able to efficiently transport the optical energy, due to polarization effects within the metals, to spatial distances farther away from the active silver surface, up to several tenth of nanometers.\\
In a concentric model, the hybrid device reveals an enhancement exceeding that of the pure Ag system by a factor of two to seven, depending on the spectral separation with respect to the Ag plasmon resonance peak. The dependence on the different layer thicknesses enables us to suggest optimal coating sizes. For example, a dielectric layer with high permittivity and a rather thin Au coating yields a stronger enhancement.\\
In the spherical model an extension to multiple coatings is straightforward. Within this approach only dipole contributions are non--vanishing and different configurations require a change in dielectric functions and radii only.\\
The image dipoles approach provides a description with non--vanishing excitation of higher multipoles for non--spherical geometry. Symmetry breaking  due to surface defects leads to higher enhancement factors at the surface of the Ag--Au hybrid compared to the Ag device. The most important difference refers to the polarization field within the planar Au layer that is not fully anti--parallel and, therefore, does not significantly attenuate the field enhancement.\\
It has been shown, that this approach predicts comparable enhancement factors at the outer surfaces of both devices in a non--concentric geometry. Our results are in agreement with recent experimental results.\cite{0_Feng08}
%
%
\begin{acknowledgments}
 Financial support by the Fonds der Chemischen Industrie and the DFG --- Cluster of Excellence UniCat --- is gratefully acknowledged.
\end{acknowledgments}
%

%
\end{document}